\documentclass[aps,prd,nofootinbib,twocolumn,superscriptaddress]{revtex4}

\usepackage{graphicx}

\newcommand{\beq}{\begin{equation}}
\newcommand{\eeq}{\end{equation}}

\newcommand{\bea}{\begin{eqnarray}}
\newcommand{\eea}{\end{eqnarray}}

\newcommand{\gsim}{\lower.7ex\hbox{$\;\stackrel{\textstyle>}{\sim}\;$}}
\newcommand{\lsim}{\lower.7ex\hbox{$\;\stackrel{\textstyle<}{\sim}\;$}}

\begin{document}

\title{Fermi-LAT gamma-ray signal from Earth Limb, systematic detector effects and their
implications for the 130~GeV gamma-ray excess
 }

\author{Andi Hektor}
\email[]{andi.hektor@cern.ch}
\affiliation{NICPB, Ravala 10, 10143 Tallinn, Estonia}
\affiliation{Helsinki Institute of Physics, P.O. Box 64, FI-00014, Helsinki, Finland}
\author{Martti Raidal}
\email[]{martti.raidal@cern.ch}
\affiliation{NICPB, Ravala 10, 10143 Tallinn, Estonia}
\affiliation{Institute of Physics, University of Tartu, Estonia}
\author{Elmo Tempel}
\email[]{elmo@aai.ee}
\affiliation{NICPB, Ravala 10, 10143 Tallinn, Estonia}
\affiliation{Tartu Observatory, Observatooriumi 1, T\~oravere 61602, Estonia}

\date{\today}

\begin{abstract}
We look for possible spectral features and systematic effects in Fermi-LAT publicly available high-energy gamma-ray data by studying photons from the Galactic centre, nearby galaxy clusters, nearby brightest galaxies, AGNs, unassociated sources, hydrogen clouds and Earth Limb. 
Apart from already known 130 GeV gamma-ray  excesses from the first two sources, we find no new statistically significant signal from others.   
Much of our effort goes to studying Earth Limb photons. In the energy range 30 GeV to 200 GeV the Earth Limb gamma-ray spectrum follows
power-law with spectral index $2.87\pm 0.04$ at 95\% CL, in a good agreement with the PAMELA measurement of cosmic ray proton spectral 
index between 2.82-2.85, confirming the physical origin of the Limb gamma-rays. In small subsets of Earth Limb data with small photon incidence 
angle it is possible to obtain spectral features at different energies, including at 130 GeV, but determination of background, thus their significances,  
has large uncertainties in those cases. We observe systematic $2\sigma$ level differences in the Earth Limb spectra of gamma-rays with small 
and large incidence angles. The behaviour of  those spectral features as well as background indicates that they are likely  statistical fluctuations.

\end{abstract}


\maketitle

\section{Introduction}
The existence of cold dark matter (DM) of the Universe~\cite{Bertone:2004pz}  is verified beyond any reasonable doubt~\cite{Komatsu:2010fb}.
The leading paradigm is that the DM consist of weakly interacting massive particles (WIMPs) whose thermal relic abundance is predicted to be 
around the observed value if WIMP mass is ${\cal O}(100)$~GeV. Today the nature and properties of the DM particles are not known. 
In comparison with other search methods for the DM, somewhat unexpectedly the indirect searches  for DM annihilations/decays in 
cosmic rays have generated lots of activity in the field during last few years~\cite{Cirelli:2010xx}. 

As a new development in DM indirect searches a 130~GeV gamma-ray 
line~\cite{Bergstrom:1988fp,Bringmann:2011ye} has been observed in Fermi-LAT~\cite{Atwood:2009ez} publicly available data first from 
the Galactic centre~\cite{Bringmann:2012vr,Weniger:2012tx,Tempel:2012ey,Su:2012ft}  and then from the six nearby galaxy clusters~\cite{Hektor:2012kc,Huang:2012yf}  consistently with Fermi-LAT constraints~\cite{Ackermann:2012qk}.  
In addition, an evidence for a double peak structure in photons from Fermi unassociated sources has been claimed~\cite{Su:2012zg}. 
However, this result has been criticized by analyzing the low energy~\cite{Hooper:2012,Mirabal:2012za}  as well as the high energy~\cite{Hektor:2012jc} 
Fermi data. Both the Galactic centre and the galaxy clusters are known to be DM dominated objects and 
the most promising known places  for searches for DM signals. 
The coincidence of the two excesses both in shape and in energy suggest that they both originate from DM annihilations 
(disfavouring DM decays due to large boost factor in galaxy clusters)~\cite{Hektor:2012kc}.
The expected signal from nearby dwarf galaxies is too weak for detection with present Fermi-LAT statistics~\cite{GeringerSameth:2012sr,Huang:2012yf}. 
The improved Fermi-LAT energy resolution and larger statistics would eventually 
allow one to tell whether the observed 130~GeV peak is a single monochromatic
gamma-ray line~\cite{Weniger:2012tx,Tempel:2012ey},  perhaps from internal brehmstrahlung~\cite{Bringmann:2012vr},
two narrow lines from $\gamma\gamma$ and $Z\gamma$ final states~\cite{Cline:2012nw,Rajaraman:2012db,Su:2012ft} or a
narrow box-like spectrum~\cite{ibarra}.
The observed spectral feature is shown  not to be consistent with broad photon spectra induced by conventional DM annihilation modes~\cite{Cirelli:2008pk} 
to charged standard model (SM) particles  or with broad box like spectrum predicted by DM annihilations 
via light intermediate states~\cite{ArkaniHamed:2008qn}.  Additional constraints
occur from soft secondary spectrum induced by $Z,H$ or other non-stable final state particles~\cite{Cohen:2012me}.
This result imposes severe constraints on building models of DM with those properties~\cite{Beacom:2004pe,Bergstrom:2004cy,Bergstrom:2005ss,Gustafsson:2007pc,Goodman:2010qn,Profumo:2010kp,Jackson:2009kg,Dudas:2009uq,Mambrini:2009ad,Ferrer:2006hy,Chalons:2011ia,
Dudas:2012pb,Choi:2012ap,Lee:2012bq,Kyae:2012vi,Acharya:2012dz,Buckley:2012ws,Chu:2012qy,Weiner:2012cb,Feng:2012gs,Das:2012ys,Kang:2012bq,Heo:2012dk,Buchmuller:2012rc,Yang:2012ha,Cholis:2012fb,Oda:2012fy,Frandsen:2012db,Park:2012xq,Tulin:2012uq,Cline:2012bz,Bai:2012qy,Bergstrom:2012uy,Wang:2012uy,Weiner:2012gm,Lee:2012wz,Baek:2012ub,Shakya:2012fj}.
Future observations are able to distinguish between those scenarios definitively~\cite{Li:2012qg,Bergstrom:2012vd,Laha:2012he,Murase:2012uy}. 

Although the observed 130~GeV gamma-ray excess in Fermi-LAT data is perfectly consistent with the DM annihilation scenario,
one should worry whether this actually is a signal of  new physics or not. It is not excluded that this excess is just an upward fluctuations
of the background~\cite{Boyarsky:2012ca} or a systematic detector effect.  Although the Fermi-LAT detector is calibrated both with Monte Carlo and with real data~\cite{Ackermann:2012kc}, and the available datasets are reconstructed accordingly, there is still a possibility that some systematic
instrumental feature  affects the photon spectrum at very high-energies. If one finds that the observation indeed represents  a true signal,
 one should still be worried whether Fermi-LAT sees some new astrophysical phenomenon~\cite{Profumo:2012tr,Aharonian:2012cs}
  or new physics beyond the SM.  Comparison of data obtained from different sources allows one to address those questions.

The aim of this work is twofold. First, we  search for and study spectral features  in gamma-rays coming from 
several known cosmological objects like nearby galaxies, dwarf galaxies, active galactic nuclei (AGN),  
molecular hydrogen clouds, compact high-velocity clouds, Earth albedo and Earth Limb in addition to the galactic centre and galaxy clusters.
If the 130~GeV feature is also seen in data from any of those places, 
 this would allow us to study and to understand the origin of the peak.
Second, we search for possible detector dependent systematic effects in the available gamma-ray data 
by comparing gamma-ray spectra from different sources. Our aim is to find out 
 whether the 130~GeV excess could show some systematic instrumental featrures.
 When  doing that we focus on the possible  dependence of the gamma-ray spectra on spatial origin,
on the incidence angle $\theta$ and the zenith angle  $Z$.
The first choice  is motivated by the fact that the 130~GeV excess is  seen only in very particular places in the sky not in all data.
The second choice is motivated by the  possible presence of  a 130~GeV spectral feature also in the Earth Limb data~\cite{wenigerlimb,Bringmann:2012mx}.  
If true, this would be a surprising result since this data is used to calibrate the Fermi LAT detector~\cite{Ackermann:2012kc} and no
spectral features are seen  by Fermi-LAT Collaboration in the early data~\cite{Abdo:2009gt}. 
Therefore a careful study of possible systematic dependences of the available gamma-ray spectra on  $\theta$ and on $Z$
is well motivated.

We find no statistically significant spectral features in the gamma-ray spectra in any of the above mentioned new sources.
Therefore we proceed to detailed studies of Earth Limb gamma-rays.
No spectral features  occur in the total Earth Limb gamma-ray data  that follows a power-law with spectral index $2.87\pm0.04$ at 95\% CL. 
This is in a perfect agreement with the PAMELA measurement of diffuse cosmic ray proton flux  at Earth with the spectral index 
2.82-2.85, depending on the considered energy range~\cite{Adriani:2011cu}, confirming the physical origin of the Limb photon flux. 

We note that the gamma-ray data from the Earth Limb is  obtained in different Fermi observation modes. Statistically dominant Earth Limb data is obtained 
during the normal survey mode of LAT and is featured by very large incidence (off-axis) angle $\theta$ and the zenith angle $110^\circ<Z<114^\circ.$
In the normal survey mode this signal is considered as a very bright  Earth atmosphere background to cosmological observations and is cut out by
requiring small values of the zenith angle.
An order of magnitude smaller fraction of Earth Limb data is collected during special observation periods when the LAT is facing Earth,
possibly observing objects close to the atmosphere. One can analyze those sets of data separately to search for possible
systematic effects.  We study an incidence angle and time dependences of the Earth Limb signal. We find that in some cases 
spectral features occur in the statistically limited gamma-ray spectra. However, in those cases also determination of the background
suffers from large fluctuations. We observe some difference in spectral indices of large and small incidence angle photons
from Earth Limb. This effect may be due to small statistics of the latter dataset since, if present, it should be a detector effect.
 We conclude that the observed trends in the statistically limited small incidence angle Earth Limb data 
 contradict trends in much larger survey mode data, and are most likely just statistical fluctuations.
 New dedicated Fermi-LAT observations of the Earth Limb could resolve this issue.

Similarly, we do not observe any obvious systematic effect that could discriminate the 130~GeV peak photons from
the background ones, in agreement with the findings of Ref.~\cite{Whiteson:2012}.
The 130~GeV gamma-ray  peak occurs in some spatial regions and does not occur in others, showing no systematic features in any
parameter. Although the observed  {\it excess} over the power-law background  cannot be explained with systematic
instrumental effects, we still can speculate that its peak-like shape might be due to a systematic {\it deficit} at 105~GeV. 
In this case the deficit must be nontrivially correlated with an increase of charged cosmic ray fluxes from the directions that
show the gamma-ray excess causing the detector effect. We find this possibility unlikely since the peak-like excess has also 
observed in a stacked data from many directions.

The paper is organized as follows. In section II we present details of our data analyses.
In section III we present our results. We conclude in section IV.

\section{Data analyses}

In the present analysis, we consider the public Fermi-LAT~\cite{Atwood:2009ez} photon event data of 210 weeks (from 4 Aug 2008 to 5 Aug 2012) within energy region from 20 to 300~GeV. We apply the recommended quality-filter cut $\mathrm{\mbox{DATA\_QUAL}}=1$, $\mathrm{\mbox{LAT\_CONFIG}}=1$. We make use of the \mbox{ULTRACLEAN} events selection (Pass~7 Version~6), in order to minimise potential systematical errors. We also tested \mbox{CLEAN} and \mbox{SOURCE} events selections, having negligible effect to our results. The selection of events was performed using the 18 April 2012 version of ScienceTools~\mbox{v9r27p1}.

For Earth Limb analyses we use the zenith-angle region $110^\circ<Z<114^\circ$: most of the Limb photons come from the region $112\pm 1^\circ$. Angle $\theta$ is the incidence angle between the photon and the axis of the telescope also called the off-axis angle: for $\theta=0^\circ$ photons go directly to the telescope. The Earth Limb photons are divided into two subsets. For the first subset, which are collected during the all-sky scanning survey, $\mathrm{ABS(ROCK\_ANGLE)}<52$.   All Earth Limb photons that come more-or-less directly to the telescope are from observations where $\mathrm{ABS(ROCK\_ANGLE)}>52^\circ$. When analyzing other regions than Earth Limb, we apply the zenith-angle cut $\theta<105^\circ$ in order to avoid contamination from the Earth photons, as recommended by the Fermi-LAT team.

To avoid the effect of point sources we exclude photons that are within an energy-independent cut radius of each source. We used all (1873) sources from the LAT 24 month catalog \cite{Nolan:2012}. The cut radius is considered $0.2^\circ$ \cite{Ackermann:2012qk}. In addition, we tested the radii 0.15$^\circ$, 0.25$^\circ$ and 0.5$^\circ$ resulting no significant effect on final results. To avoid contamination with Galactic plane, we exclude all photons with $|b|<5^\circ$.

We use kernel smoothing method to fit the gamma-ray data. This method should be optimal for searching for peak-like spectral features.
For technical details we refer the reader to Ref.~\cite{Tempel:2012ey}. However, we note that we have updated our analyses compared to 
Ref.~\cite{Tempel:2012ey}  using new improved Fermi-LAT energy resolution and calibrated data taking into account angular dependences 
presented in Ref.~\cite{Ackermann:2012kc}.
The errors to the fitted spectra as well as to the power-law fits are calculated with bootstrap.
All errors in this work are at 95\% CL ($2\sigma$).

\section{Gamma-ray spectra, possible systematics and correlations}

\subsection{Search for spectral features from various sources and systematics}

\begin{table}
    \caption{Regions in the Galaxy featuring large number of $20<E_\gamma< 300$ GeV photons without 130 GeV excess.  }
    \begin{tabular}{clclclcl}
        \hline\hline
        Region &  $l$ (deg)  &  $b$ (deg)  & $N_\gamma$  \\
        \hline
        1  &  18.0  &  0.0 &  946 \\
        2  &  25.0  &  0.0 &  943 \\
        3  &  331.0  &  0.0 &  1041 \\
        4  &  337.0  &  0.0 &  1137 \\
        5  &  346.0  &  0.0 &  933 \\
          \hline
    \end{tabular}
\label{regions}
\end{table}

We searched for the 130~GeV peak in gamma-rays from several sources: from nearby bright galaxies (extracted from 2MASS galaxy redshift catalogue \cite{Lavaux:11}), dwarf galaxies, bright AGNs \cite{Abdo:09}, and from compact high-velocity clouds \cite{deHeij:02,Putman:02}. From all these sources we stacked the signal as we did in the case of studying the signal from galaxy clusters \cite{Hektor:2012kc}. Additionally, we looked the signal from Andromeda and Crab nebula. None of these sources yield statistically significant signal  of the 130~GeV gamma ray line. However, for completeness we note
that  in the case of  nearby molecular clouds, recently studied by Fermi-LAT Collaboration~\cite{Ackermann:2012mt}, we do observe    
a peak-like feature at 130~GeV with statistical significance less than $2\sigma.$ Most likely this is nothing but a statistical 
fluctuation. 

\begin{figure}[t]
\begin{center}
\includegraphics[width=0.5\textwidth]{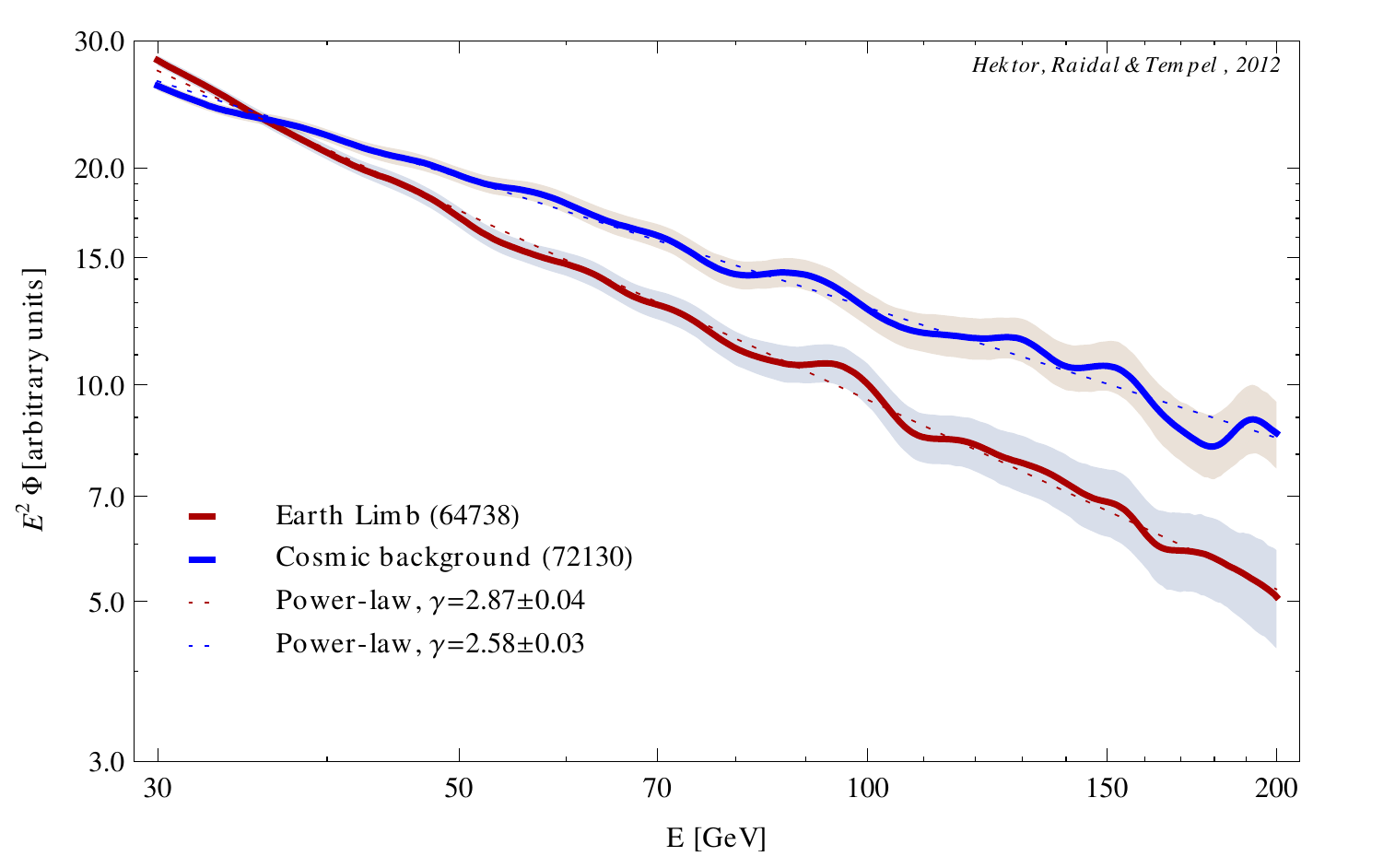}
\caption{
Spectra of Fermi-LAT gamma-rays from the cosmic background ($\theta<62^\circ,$ $Z<105^\circ$)  and  
from the Earth Limb ($110^\circ<Z<114^\circ$),  together with $2\sigma$ error bands, as functions of photon energy.
The fluxes are in arbitrary units, the corresponding numbers of photons are presented in the figure. 
Best power-law fits to data together with their spectral indices are also presented.
 } 
\label{fig1}
\end{center}
\end{figure}

So far the 130~GeV excess has been claimed from the Galactic centre, nearby galaxy clusters, from several small regions in the Galaxy
listed in Ref.~\cite{Tempel:2012ey}  and from some unassociated Fermi sources~\cite{Su:2012zg}. It is tempting to associate the last two classes of 
sources with DM subhaloes. However, those may also be just statistical fluctuations.
 At the same time, the overall integrated cosmic gamma-ray spectrum defined by $\theta<62^\circ,$ $Z<105^\circ$, 
 presented in Fig.~\ref{fig1} with a blue line, does not show any excess at 130~GeV. 
 It follows a power-law spectrum with spectral index $2.58\pm 0.03$  and appears as a background to the observed 130 GeV excesses. 
 The $2\sigma$ error band is also presented in   Fig.~\ref{fig1}  with a grey band. 
 In addition,  there are several bright region in the sky, some of them are listed in Table~\ref{regions} for comparison with the similar regions listed 
in Ref.~\cite{Tempel:2012ey}, that do not show any 130 GeV excess in the spectrum.
Comparing results from all those places does not allow us to propose any logically consistent way to explain the 130~GeV peak
with systematic  detector effect. First, it is difficult to explain why the detector effect must give an excess. Second, we see no correlations of the
the excess with any particular direction, with the the incidence angle nor with the number of photons from the signal region. 
The 130 GeV line seems to be a real excess.  

Although explaining the 130 GeV excess over the power-law background due to systematic detector effects seems to be 
disfavoured, one can still assume that the excess is due to new source of high-energy gamma-rays but its peak-like shape is 
a detector effect. This assumption implies that the excess is broad but there is a systematic deficit around photon energies 
105 GeV that produces peak-like spectrum. For DM scenarios this scenario would imply that most of DM annihilation
modes that induce broad spectrum can explain the excess. However, similarly to the previous case, this scenario is also 
disfavoured by non-observation of 105 GeV deficit in data from other sources. However, such an effect might be correlated with 
charged cosmic ray fluxes and  requires dedicated studies.

\begin{figure}[t]
\begin{center}
\includegraphics[width=0.47\textwidth]{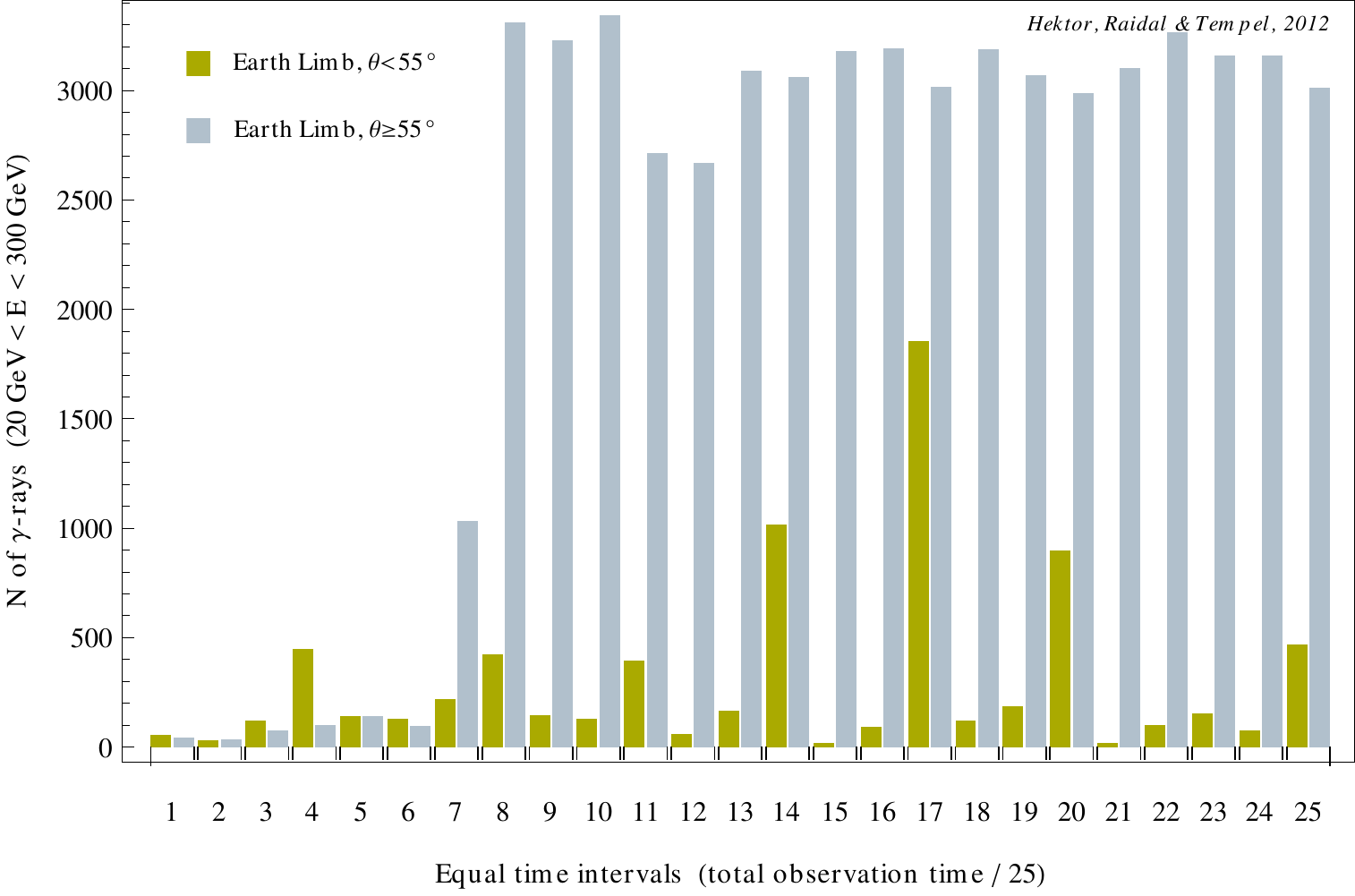}
\caption{
Earth Limb data collected by Fermi-LAT as a function of time. The grey (light) 
bars correspond to $N=64738$ photons with large incidence angle $\theta>55^\circ$  
collected during the normal cosmic survey mode. The green (dark)  bars correspond to 
$N=7467$ small incidence angle photons collected when the detector faces Earth. 
 } 
\label{fig2}
\end{center}
\end{figure}

\subsection{Limb photons and possible systematics}

Earth atmosphere is a bright source of high-energy gamma-rays that is a background for cosmic observations. Therefore 
in the cosmic observation mode the detector faces away from the Earth. Above $E_\gamma \gtrsim 1$ GeV, the brightest region in the Earth atmosphere is its Limb in which the diffuse cosmic rays, dominated by protons, collide  with the outer layer of the atmosphere producing 
gamma-ray flux. The Fermi-LAT sees the photons which are directed to the detector. Moving in the direction of the 
Earth itself the atmosphere starts to absorb all cosmic rays and the Earth albedo has only a soft photon spectrum.
Because of this production mechanism, the Earth Limb signal in gamma-rays is expected have approximately the same 
spectral index as the diffuse proton flux at the Earth. The latter is measured with very high precision by PAMELA satellite yielding 
the following values of spectral index>\cite{Adriani:2011cu},
\bea
\gamma_p &=& 2.820\pm 0.004 \;\;\;{\rm for} \;\; 30~\rm{GeV} <E<1.2~ \rm{TeV}, \nonumber \\
\gamma_p &=& 2.850\pm 0.016 \;\;\;{\rm for} \;\; 80~\rm{GeV} <E<230~ \rm{GeV},
\label{PAMELA}
\eea
depending on the proton energy ranges.

In this work we consider Earth Limb to correspond to the zenith angles $110^\circ<Z<114^\circ$.
In fact most of the Limb photons come from the region $Z=112\pm 1^\circ$. 
We plot in Fig.~\ref{fig1} the gamma-ray flux from Earth Limb as a function of the photon energy
together with $2\sigma$ error band. The Earth Limb data follows a featureless power law with spectral index $2.87\pm 0.04$.
This is in a good agreement with Eq.~(\ref{PAMELA}). Since the two experiments, Fermi LAT and PAMELA,
measure the same spectral index for the Earth Limb gamma-ray and for the proton fluxes, respectively,
we conclude that the production mechanism for Earth Limb gamma-rays is in good agreement with theoretical prediction.
Notice also that both the cosmic background and Earth Limb spectra in Fig.~\ref{fig1}  are featureless, both datasets contain approximately
the same number of photons, and their spectral indices are clearly different. Thus the physical origin of the two spectra are different.

\begin{figure}[t]
\begin{center}
\includegraphics[width=0.5\textwidth]{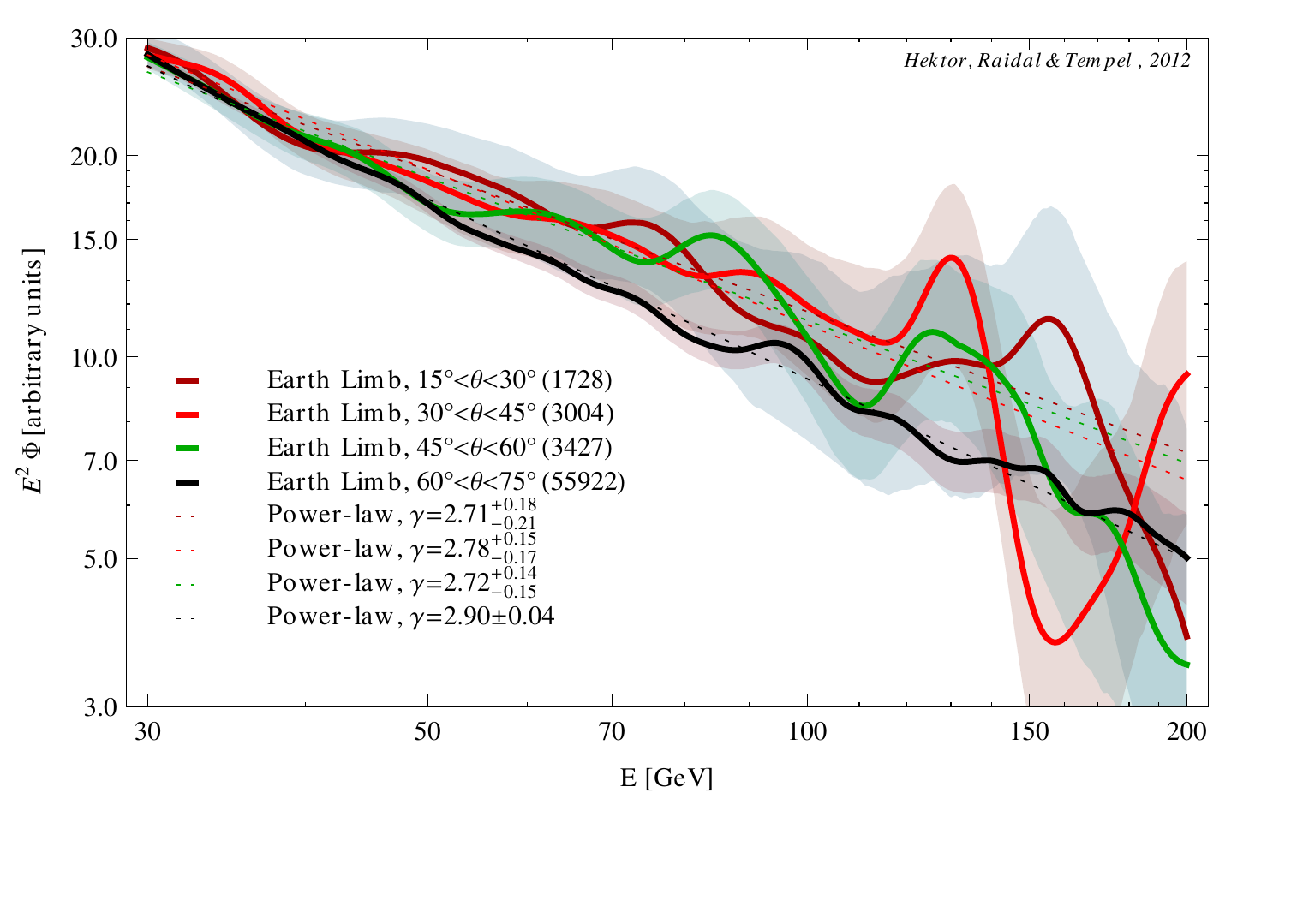}
\caption{
Spectra of the Earth Limb photons divided into four datasets by the incidence angle as functions of energy.
The chosen incidence angle regions, the numbers of photons in those and best fit power-law spectral
indices  are all presented in the figure. The shaded areas show $2\sigma$ error-bands calculated with bootstrap technique. 
 } 
\label{fig3}
\end{center}
\end{figure}

During the normal Fermi-LAT cosmological survey mode  ($\mathrm{rock\_angle}<52^\circ$) the detector is facing away from the Earth and 
Earth Limb photons arrive to the detector only at very large incidence angles. At the same time there 
exist Limb data collected when detector faces the Earth atmosphere,  $\mathrm{rock\_angle}>52^\circ$. 
To demonstrate how the Limb data is
collected over the time we present in Fig.~\ref{fig2} the time histogram of the collected number of photons 
divided by the incidence angle. Photons with $\theta>55^\circ$ vastly dominate the data ($N=64738$)
while the number of photons collected when LAT faces Earth, $\theta<55^\circ$, is an order of magnitude smaller, $N=7467$.
The Earth Limb spectrum in Fig.~\ref{fig1}
is entirely dominated by the large incidence angle photons. In the following we study whether the spectrum depends on the
incidence angle or not.

We divide the Earth Limb photons into four datasets by the incidence angle and plot the corresponding
spectra against the photon energy in Fig.~\ref{fig3}. Normalizations are arbitrary, the number of photons and their best fit
power-law spectra are also presented.  The shaded areas show $2\sigma$ error-bands calculated with bootstrap technique
as described in~\cite{Tempel:2012ey}.
Clearly the statistically dominant datasets corresponding to $60^\circ<\theta<75^\circ$
agrees well with the total spectrum in Fig.~\ref{fig1}. However, the spectra for small incidence angle 
show spectral features at high energies. In particular, the $15^\circ<\theta<30^\circ$ dataset shows a peak at 150~GeV,
the $45^\circ<\theta<60^\circ$ dataset shows a deficit at 105~GeV  and
 the $30^\circ<\theta<45^\circ$ dataset shows a peak at 130~GeV. The existence of the latter peak has been
 used by several people anonymously (by our referees)  and publicly (private discussions) 
 to argue that its existence may indicate that the 130~GeV peak from the Galactic centre cannot be physical.  
Notice also that the spectral index of the best power-law fit to Earth Limb data with small $\theta$ is systematically 
smaller than $2.87\pm 0.04$  that is the spectral index for the large incidence angle Earth Limb data.  
This affects calculations of the statistical significances of the spectral features in Fig.~\ref{fig3}. 
Clearly, the 130~GeV peak (red line) is statistically insignificant ($\sim 2.7\sigma$) when compared with the
power-law fit to $30^\circ<\theta<45^\circ$ data (spectral index $\gamma=2.78^{+0.15}_{-0.17}$) 
 but looks much more significant ($\sim 3.5 \sigma$) over the fit to $60^\circ<\theta<75^\circ$ data with $\gamma=2.90\pm 0.04.$
However, the errors are large because the datasets are small, and the fits in Fig.~\ref{fig3} 
are consistent with each other at $2\sigma$ level.

\begin{figure}[t]
\begin{center}
\includegraphics[width=0.5\textwidth]{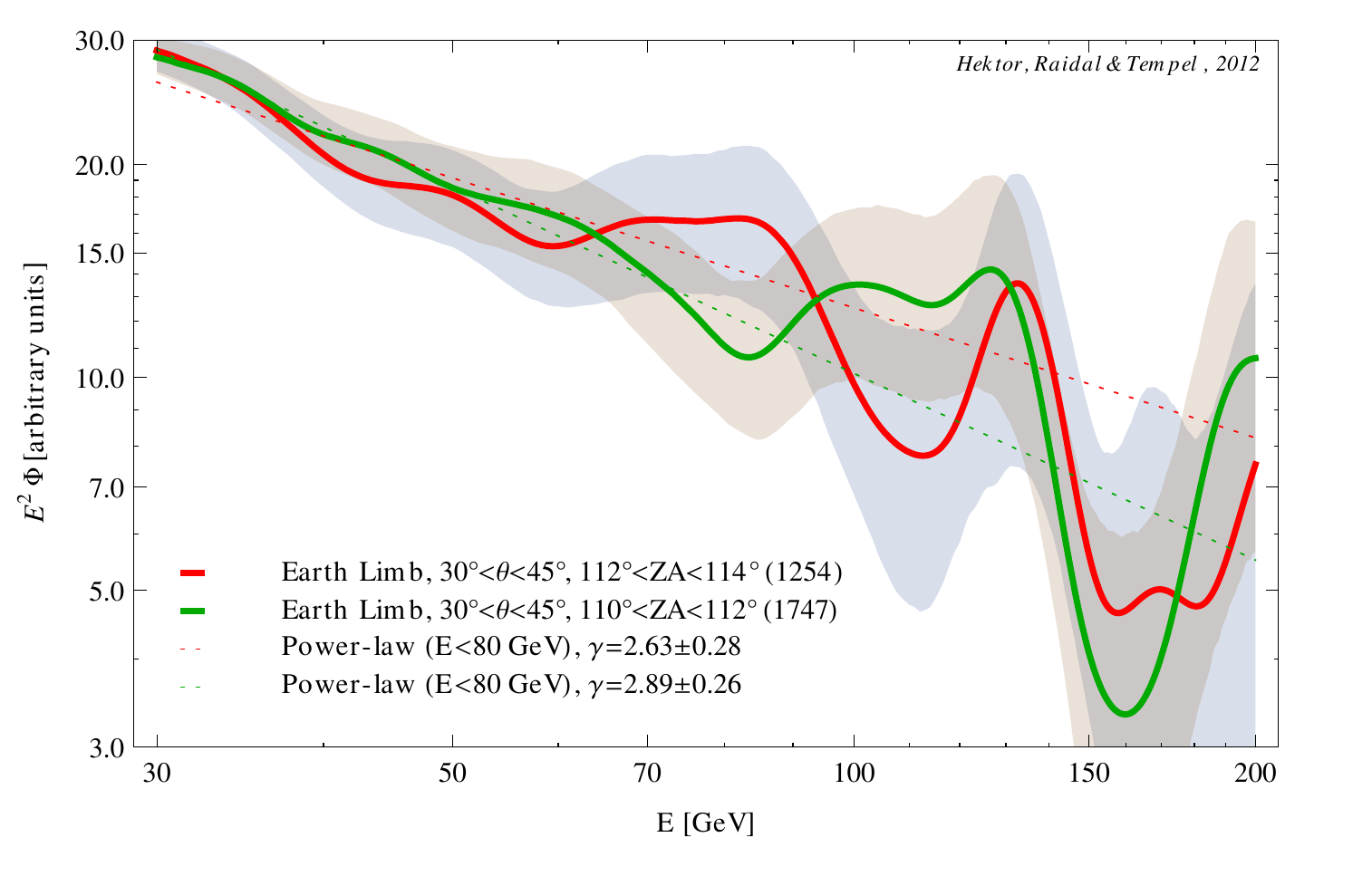}
\caption{
Gamma-ray spectra for incidence angles $30^\circ<\theta<45^\circ$ for two zenit angle ranges
$110^\circ<Z<112^\circ$ and $112^\circ<Z<114^\circ$ together with their best power-law fits. 
The latter are for photon energies $E<80$~GeV.
 } 
\label{fig4}
\end{center}
\end{figure}

Because the statistics is poor for $\theta<60^\circ$ and because there is no clear
systematic feature shared by all the lines in Fig.~\ref{fig3}, our first explanations to those findings is statistical fluctuation.
To study those features further we concentrate on the incidence angle region $30^\circ<\theta<45^\circ$
and split the gamma-rays  from that region into two datasets according to their origin from either from the inner or the outer
half of the Earth Limb corresponding to $110^\circ<Z<112^\circ$ and $112^\circ<Z<114^\circ$, respectively.
The results are presented in Fig.~\ref{fig4}. The two spectra have opposite behaviour at 80 GeV and 
different shapes of excesses between 100-130~GeV. At larger photon energies the fluctuations become large due to the lack of
statistics. Assuming that up to the photon energies $E<80$~GeV the statistics is sufficient for determining the power-law fit to data,
we also present those fits in the figure. The spectral indices of those are quite different but consistent within errors. Because no systematic
difference in the behaviour of the Limb photons is expected for $110^\circ<Z<112^\circ$ and for $112^\circ<Z<114^\circ$,
those results indicate fluctuations due to limited statistics.

\begin{figure}[t]
\begin{center}
\includegraphics[width=0.5\textwidth]{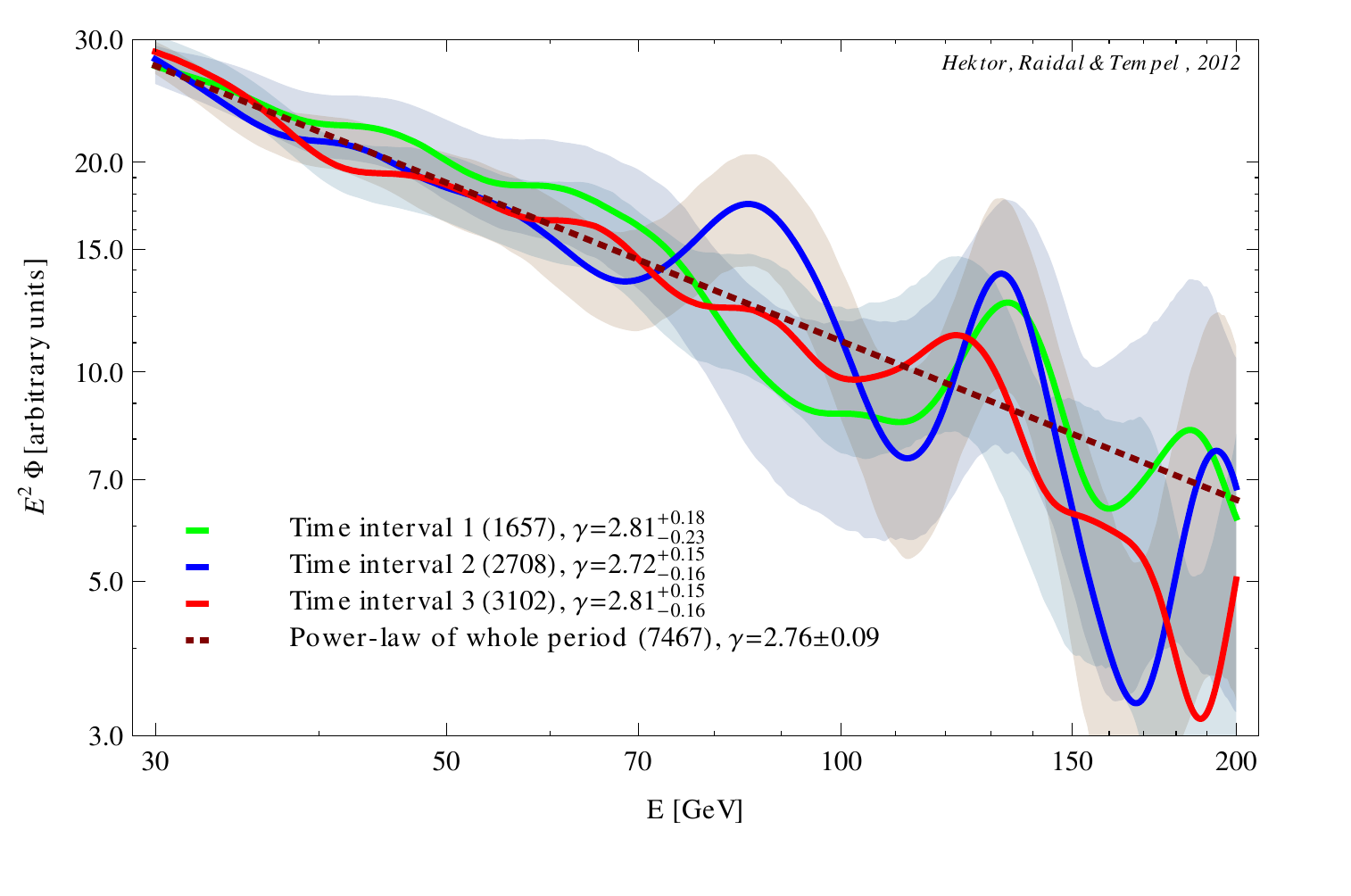}
\caption{
Spectra of Limb photons with small incidence angles, $\theta<55^\circ$, collected by Fermi-LAT during three equal time intervals.
The best power-law fit to all $\theta<55^\circ$ Earth Limb data is shown in dotted line. 
 } 
\label{fig5}
\end{center}
\end{figure}

To study the time dependence of the spectral features we split the Earth Limb data with small incidence angles,
$\theta<55^\circ$, presented with green bars in Fig.~\ref{fig2}, into three datasets according to the time the events are recorded. 
The time intervals are all equal but the number of recorder photons depends on the time interval. 
The resulting spectra are plotted in Fig.~\ref{fig5}. In addition we plot in Fig.~\ref{fig5} the best power-law fit 
to the total Earth Limb data with  $\theta<55^\circ$. The resulting spectral index is 
$\gamma = 2.76 \pm 0.09$ that should be compared with the spectral index of the fit to 
all Earth Limb data  $\gamma = 2.87\pm 0.04$ plotted in Fig.~\ref{fig1}. The two agree within
$2\sigma$ errors. Such a systematic dependence of the power-law spectral index on the incidence angle 
is qualitatively observed by Fermi and we used the results presented in Ref.~\cite{Ackermann:2012kc} to calibrate data. 
However, we find that this is quantitatively just a small effect that cannot explain our finding. 
The trend  that the small and the large incidence angle 
gamma-rays tend to follow somewhat different power-law is a possible systematic feature in the data.

\section{Discussion}

The 130~GeV gamma-ray excess from the Galactic centre, that is statistically the most significant one so far, is visible only in photons coming to
the detector with small incidence angles.  We took the photons from Galaxy centre with $\theta>55^\circ$ and we do not observe 
any peak signal in this case.  
However, the number of photons for this data sample is relatively small and we cannot draw any statistically meaningful conclusion.
Studies of the Earth Limb photons in this work show the same behaviour: the gamma-rays coming to the detector at small off-axis angle may
show possible spectral features demonstrated in Figs.~\ref{fig3}-\ref{fig5} while the gamma-rays coming to the detector at large off-axis angle
do not show any spectral features, see Figs.~\ref{fig1} and \ref{fig3}. However, the crucial difference between those two cases is that for the Galactic centre signal the $\theta<55^\circ$ data is statistically absolutely dominant while for the Earth Limb signal the data is statistically dominated by $\theta>55^\circ$
photons. The total Earth Limb signal follows featureless power-law which spectral index 2.87 that agrees very well with the PAMELA measurement of the
diffuse proton spectral index at Earth. We believe that this coincidence is not an accident but a confirmation of the theoretical prediction that the Earth Limb
photon spectrum must have the same spectral index as the dominant charged cosmic ray background. Therefore, the Earth Limb data
confirms the known physics.

Small fraction of Earth Limb data is collected when Fermi-LAT has observed objects close to the Earth atmosphere. 
In this case photons arrive to the detector at small incidence angle. We studied this subset of data in order to find
spectral features and systematic effects in the data. As seen in Figs.~\ref{fig3}-\ref{fig5}, indeed, there are spectral features at
large energies. However, those spectral features are peaked at different energies and change their shapes when different 
subsamples of this data are considered. The possible explanations to those findings  include:
\begin{enumerate}
\item The observed features in Figs.~\ref{fig3}-\ref{fig5} are due to statistical fluctuations. The studied subsamples of data are statistically
limited, there are just few photons at large energies and one expects upward and downward fluctuations to appear. The described behaviour
of the spectral features in Figs.~\ref{fig3}-\ref{fig5} supports this explanation.

\item The small incidence angle data is collected when observing objects close to the Earth atmosphere. It is possible that those observations
are contaminated by photons coming either from those or other cosmological objects or by diffuse gamma-rays from the background. 
 We do not    observe any distinct bright regions in the Limb signal  and disfavour this explanation.

\item The spectral features could be systematic detector effects due to charged cosmic rays.  
For example, if the charged cosmic ray flux from some particular direction is higher than the usual background,
the detector may systematically not register incoming photons. For example, one could argue that there 
is a deficit of gamma-rays with 105~GeV energy. However, this would imply that  there are preferred directions 
in the diffuse proton flux at Earth that radiate more cosmic rays than others. We are not aware of those effects. 

\end{enumerate}
Our favourite explanation is the first one.

\section{Conclusions}

We have searched for the 130~GeV gamma-ray peak in photons from several nearby cosmological objects and found no new candidates
in addition to the known ones. We also searched for possible systematic effects in the 130~GeV photons and could not identify any.
The 130~GeV gamma-ray signal seems to be a real excess over the power-law background. We studied gamma-rays from
Earth Limb most extensively and found that the Earth  Limb signal confirms the known physics - the flux from the Limb follows featureless power-law with
the spectral index equal to the one measured by PAMELA for the diffuse proton flux. This is approximately the theoretical prediction for the
Earth Limb gamma-rays. We have studied in detail the possible presence of spectral features in subsamples of Earth Limb data featured by
small photon incidence angles. The behaviour of those spectra indicates that the spectral features in our  Figs.~\ref{fig3}-\ref{fig5} 
are most likely statistical fluctuations due to small number of photons in those datasets.

An independent analyses of Earth Limb data with similar conclusions on the 130~GeV peak as a systematic effect
will appear simultaneously with this paper~\cite{FSW}.

\vspace{0.3cm}

{\bf Acknowledgement.}
We thank Marco Cirelli and Gert H\"utsi for numerous discussions, and Doug Finkbeiner and Christoph Weniger for communications related to this work.
This work was supported by the ESF grants 8090, 8499, 8943, MTT8, MTT59, MTT60, MJD52, MJD272, by the recurrent financing projects SF0690030s09, SF0060067s08 and by the European Union through the European Regional Development Fund.


\begin{thebibliography}{99}

  
   
\bibitem{Bertone:2004pz} 
For a review see,   G.~Bertone, D.~Hooper and J.~Silk,
  Phys.\ Rept.\  {\bf 405}, 279 (2005)
  [hep-ph/0404175].
  
  
\bibitem{Komatsu:2010fb}
  E.~Komatsu {\it et al.}  [WMAP Collaboration],
  ``Seven-Year Wilkinson Microwave Anisotropy Probe (WMAP) Observations: Cosmological Interpretation,''
  Astrophys.\ J.\ Suppl.\  {\bf 192} (2011) 18
  [arXiv:1001.4538 [astro-ph.CO]].

  
\bibitem{Cirelli:2010xx} 
  For constantly updated review see, 
  M.~Cirelli, G.~Corcella, A.~Hektor, G.~Hutsi, M.~Kadastik, P.~Panci, M.~Raidal and F.~Sala {\it et al.},
  JCAP {\bf 1103}, 051 (2011)
  [arXiv:1012.4515 [hep-ph]].
  http://www.marcocirelli.net/PPPC4DMID.html
  
  
\bibitem{Bergstrom:1988fp}
  L.~Bergstrom and H.~Snellman,
  ``Observable Monochromatic Photons From Cosmic Photino Annihilation,''
  Phys.\ Rev.\ D {\bf 37} (1988) 3737.
  
\bibitem{Bringmann:2011ye}
For complete set of references see, \\
  T.~Bringmann, F.~Calore, G.~Vertongen and C.~Weniger,
  ``On the Relevance of Sharp Gamma-Ray Features for Indirect Dark Matter Searches,''
  Phys.\ Rev.\ D {\bf 84} (2011) 103525
  [arXiv:1106.1874 [hep-ph]].

 
\bibitem{Atwood:2009ez}
  W.~B.~Atwood {\it et al.}  [LAT Collaboration],
  Astrophys.\ J.\  {\bf 697} (2009) 1071
  [arXiv:0902.1089 [astro-ph.IM]].


\bibitem{Bringmann:2012vr} 
  T.~Bringmann, X.~Huang, A.~Ibarra, S.~Vogl and C.~Weniger,
  arXiv:1203.1312 [hep-ph].
  
\bibitem{Weniger:2012tx} 
  C.~Weniger,
  arXiv:1204.2797 [hep-ph].

\bibitem{Tempel:2012ey} 
  E.~Tempel, A.~Hektor and M.~Raidal,
  arXiv:1205.1045 [hep-ph].

\bibitem{Su:2012ft} 
  M.~Su and D.~P.~Finkbeiner,
  arXiv:1206.1616 [astro-ph.HE].

\bibitem{Hektor:2012kc}
  A.~Hektor, M.~Raidal and E.~Tempel,
  arXiv:1207.4466 [astro-ph.HE].

\bibitem{Huang:2012yf}
  X.~-Y.~Huang, Q.~Yuan, P.~-F.~Yin, X.~-J.~Bi and X.~-L.~Chen,
  arXiv:1208.0267 [astro-ph.HE].



\bibitem{Ackermann:2012qk} 
  M.~Ackermann {\it et al.}  [LAT Collaboration],
  arXiv:1205.2739 [astro-ph.HE].

\bibitem{GeringerSameth:2012sr}
  A.~Geringer-Sameth and S.~M.~Koushiappas,
  arXiv:1206.0796 [astro-ph.HE].


\bibitem{Cline:2012nw} 
  J.~M.~Cline,
  arXiv:1205.2688 [hep-ph].
  
\bibitem{Rajaraman:2012db}
  A.~Rajaraman, T.~M.~P.~Tait and D.~Whiteson,
  arXiv:1205.4723 [hep-ph].
  
  
  \bibitem{ibarra}
  A.~Ibarra, S.~Gehler and M.~Pato,
  arXiv:1205.0007 [hep-ph].
  
  
  
\bibitem{Su:2012zg} 
  M.~Su and D.~P.~Finkbeiner,
  arXiv:1207.7060 [astro-ph.HE].
  
\bibitem{Hooper:2012} 
  D.~Hooper and T.~Linden,
  arXiv:1208.0828 [hep-ph].


\bibitem{Mirabal:2012za}
  N.~Mirabal,
  arXiv:1208.1693 [astro-ph.HE].
  
\bibitem{Hektor:2012jc}
  A.~Hektor, M.~Raidal and E.~Tempel,
  arXiv:1208.1996 [astro-ph.HE].
  

  
  
\bibitem{Cirelli:2008pk}
  M.~Cirelli, M.~Kadastik, M.~Raidal and A.~Strumia,
  Nucl.\ Phys.\ B {\bf 813} (2009) 1
  [arXiv:0809.2409 [hep-ph]].
  
\bibitem{ArkaniHamed:2008qn}
  N.~Arkani-Hamed, D.~P.~Finkbeiner, T.~R.~Slatyer and N.~Weiner,
  Phys.\ Rev.\ D {\bf 79} (2009) 015014
  [arXiv:0810.0713 [hep-ph]].

  
  \bibitem{Cohen:2012me}
  T.~Cohen, M.~Lisanti, T.~R.~Slatyer and J.~G.~Wacker,
  arXiv:1207.0800 [hep-ph].
  

\bibitem{Beacom:2004pe}
  J.~F.~Beacom, N.~F.~Bell and G.~Bertone,
  Phys.\ Rev.\ Lett.\  {\bf 94} (2005) 171301
  [astro-ph/0409403].

\bibitem{Bergstrom:2004cy}
  L.~Bergstrom, T.~Bringmann, M.~Eriksson and M.~Gustafsson,
  Phys.\ Rev.\ Lett.\  {\bf 94} (2005) 131301
  [astro-ph/0410359].

\bibitem{Bergstrom:2005ss}
  L.~Bergstrom, T.~Bringmann, M.~Eriksson and M.~Gustafsson,
  Phys.\ Rev.\ Lett.\  {\bf 95} (2005) 241301
  [hep-ph/0507229].
  
 
\bibitem{Gustafsson:2007pc}
  M.~Gustafsson, E.~Lundstrom, L.~Bergstrom and J.~Edsjo,
  Phys.\ Rev.\ Lett.\  {\bf 99} (2007) 041301
  [astro-ph/0703512 [ASTRO-PH]].
  
\bibitem{Goodman:2010qn}
  J.~Goodman, M.~Ibe, A.~Rajaraman, W.~Shepherd, T.~M.~P.~Tait and H.~-B.~Yu,
  Nucl.\ Phys.\ B {\bf 844} (2011) 55
  [arXiv:1009.0008 [hep-ph]].
 
\bibitem{Profumo:2010kp}
  S.~Profumo, L.~Ubaldi and C.~Wainwright,
  Phys.\ Rev.\ D {\bf 82} (2010) 123514
  [arXiv:1009.5377 [hep-ph]].
  
\bibitem{Jackson:2009kg}
  C.~B.~Jackson, G.~Servant, G.~Shaughnessy, T.~M.~P.~Tait and M.~Taoso,
  JCAP {\bf 1004} (2010) 004
  [arXiv:0912.0004 [hep-ph]].
  
\bibitem{Dudas:2009uq}
  E.~Dudas, Y.~Mambrini, S.~Pokorski and A.~Romagnoni,
  JHEP {\bf 0908} (2009) 014
  [arXiv:0904.1745 [hep-ph]].
  
\bibitem{Mambrini:2009ad}
  Y.~Mambrini,
  JCAP {\bf 0912} (2009) 005
  [arXiv:0907.2918 [hep-ph]].
  
\bibitem{Ferrer:2006hy}
  F.~Ferrer, L.~M.~Krauss and S.~Profumo,
  Phys.\ Rev.\ D {\bf 74} (2006) 115007,
  
\bibitem{Chalons:2011ia}
  G.~Chalons and A.~Semenov,
  JHEP {\bf 1112} (2011) 055, 
  


\bibitem{Dudas:2012pb}
  E.~Dudas, Y.~Mambrini, S.~Pokorski and A.~Romagnoni,
  arXiv:1205.1520 [hep-ph].

\bibitem{Choi:2012ap}
  K.~-Y.~Choi and O.~Seto,
  arXiv:1205.3276 [hep-ph].

\bibitem{Lee:2012bq}
  H.~M.~Lee, M.~Park and W.~-I.~Park,
  arXiv:1205.4675 [hep-ph].

\bibitem{Kyae:2012vi}
  B.~Kyae and J.~-C.~Park,
  arXiv:1205.4151 [hep-ph].

\bibitem{Acharya:2012dz}
  B.~S.~Acharya, G.~Kane, P.~Kumar, R.~Lu and B.~Zheng,
  arXiv:1205.5789 [hep-ph].

\bibitem{Buckley:2012ws}
  M.~R.~Buckley and D.~Hooper,
  arXiv:1205.6811 [hep-ph].

\bibitem{Chu:2012qy}
  X.~Chu, T.~Hambye, T.~Scarna and M.~H.~G.~Tytgat,
  arXiv:1206.2279 [hep-ph].

\bibitem{Weiner:2012cb} 
  N.~Weiner and I.~Yavin,
  arXiv:1206.2910 [hep-ph].
  
\bibitem{Feng:2012gs} 
  L.~Feng, Q.~Yuan and Y.~-Z.~Fan,
  arXiv:1206.4758 [astro-ph.HE].
  
\bibitem{Das:2012ys}
  D.~Das, U.~Ellwanger and P.~Mitropoulos,
  arXiv:1206.2639 [hep-ph].

\bibitem{Kang:2012bq}
  Z.~Kang, T.~Li, J.~Li and Y.~Liu,
  arXiv:1206.2863 [hep-ph].

\bibitem{Heo:2012dk}
  J.~H.~Heo and C.~S.~Kim,
  arXiv:1207.1341 [astro-ph.HE].

\bibitem{Buchmuller:2012rc}
  W.~Buchmuller and M.~Garny,
  arXiv:1206.7056 [hep-ph].



 
\bibitem{Yang:2012ha} 
  R.~-Z.~Yang, Q.~Yuan, L.~Feng, Y.~-Z.~Fan and J.~Chang,
  arXiv:1207.1621 [astro-ph.CO].
  
\bibitem{Cholis:2012fb} 
  I.~Cholis, M.~Tavakoli and P.~Ullio,
  arXiv:1207.1468 [hep-ph].
  
\bibitem{Oda:2012fy} 
  I.~Oda,
  arXiv:1207.1537 [hep-ph].

\bibitem{Frandsen:2012db} 
  M.~T.~Frandsen, U.~Haisch, F.~Kahlhoefer, P.~Mertsch and K.~Schmidt-Hoberg,
  arXiv:1207.3971 [hep-ph].
  
\bibitem{Park:2012xq}
  J.~-C.~Park and S.~C.~Park,
  arXiv:1207.4981 [hep-ph].


\bibitem{Tulin:2012uq}
  S.~Tulin, H.~-B.~Yu and K.~M.~Zurek,
  arXiv:1208.0009 [hep-ph].
  
\bibitem{Cline:2012bz}
  J.~M.~Cline, A.~R.~Frey and G.~D.~Moore,
  arXiv:1208.2685 [hep-ph].
  
\bibitem{Bai:2012qy}
  Y.~Bai and J.~Shelton,
  arXiv:1208.4100 [hep-ph].

\bibitem{Bergstrom:2012uy}
  L.~Bergstrom,
  arXiv:1208.6082 [hep-ph].

\bibitem{Wang:2012uy}
  L.~Wang and X.~-F.~Han,
  arXiv:1209.0376 [hep-ph].

\bibitem{Weiner:2012gm}
  N.~Weiner and I.~Yavin,
  arXiv:1209.1093 [hep-ph].
  
\bibitem{Lee:2012wz}
  H.~M.~Lee, M.~Park and W.~-I.~Park,
  arXiv:1209.1955 [hep-ph].
  
\bibitem{Baek:2012ub}
  S.~Baek, P.~Ko and E.~Senaha,
  arXiv:1209.1685 [hep-ph].
  
\bibitem{Shakya:2012fj}
  B.~Shakya,
  arXiv:1209.2427 [hep-ph].
  



\bibitem{Li:2012qg} 
  Y.~Li and Q.~Yuan,
  arXiv:1206.2241 [astro-ph.HE].

\bibitem{Bergstrom:2012vd}
  L.~Bergstrom, G.~Bertone, J.~Conrad, C.~Farnier and C.~Weniger,
  arXiv:1207.6773 [hep-ph].

\bibitem{Laha:2012he}
  R.~Laha, K.~C.~Y.~Ng, B.~Dasgupta and S.~Horiuchi,
  arXiv:1208.5488 [astro-ph.CO].
  
\bibitem{Murase:2012uy}
  K.~Murase and J.~F.~Beacom,
  arXiv:1209.0225 [astro-ph.HE].
  
\bibitem{Boyarsky:2012ca} 
  A.~Boyarsky, D.~Malyshev and O.~Ruchayskiy,
  arXiv:1205.4700 [astro-ph.HE].

\bibitem{Ackermann:2012kc}
  M.~Ackermann {\it et al.}   [Fermi-LAT Collaboration],
  arXiv:1206.1896 [astro-ph.IM].


\bibitem{Profumo:2012tr} 
  S.~Profumo and T.~Linden,
  arXiv:1204.6047 [astro-ph.HE].

\bibitem{Aharonian:2012cs} 
  F.~Aharonian, D.~Khangulyan and D.~Malyshev,
  arXiv:1207.0458 [astro-ph.HE].


\bibitem{wenigerlimb}
C.~Weniger, talk given in {\it Gamma 2012}, Heidelberg. 

\bibitem{Bringmann:2012mx}
  T.~Bringmann and C.~Weniger,
  arXiv:1208.5481 [hep-ph].


\bibitem{Abdo:2009gt}
  A.~A.~Abdo {\it et al.}  [Fermi-LAT Collaboration],
  Phys.\ Rev.\ D {\bf 80} (2009) 122004
  [arXiv:0912.1868 [astro-ph.HE]].


\bibitem{Adriani:2011cu}
  O.~Adriani {\it et al.}  [PAMELA Collaboration],
  Science {\bf 332} (2011) 69
  [arXiv:1103.4055 [astro-ph.HE]].
  
  
\bibitem{Whiteson:2012}
  D.~Whiteson,
  arXiv:1208.3677 [astro-ph.HE].
  



  
    
 

\bibitem{Nolan:2012}
  [Fermi-LAT Collaboration],
  Astrophys.\ J.\ Suppl.\  {\bf 199} (2012) 31
  [arXiv:1108.1435 [astro-ph.HE]].
  

  
\bibitem{Lavaux:11}
  G.~Lavaux and M.~J.~Hudson,
  Mon.\ Not.\ Roy.\ Astron.\ Soc.\  {\bf 416} (2011) 2840
  [arXiv:1105.6107 [astro-ph.CO]].
  
\bibitem{Abdo:09}
  A.~A.~Abdo {\it et al.}  [Fermi LAT Collaboration],
  Astrophys.\ J.\  {\bf 700} (2009) 597
  [arXiv:0902.1559 [astro-ph.HE]].
 
\bibitem{deHeij:02}
  V.~de Heij, R.~Braun and W.~B.~Burton,
  astro-ph/0201249.
  
  
\bibitem{Putman:02}
  M.~E.~Putman, V.~de Heij, L.~Staveley-Smith, R.~Braun, K.~C.~Freeman, B.~K.~Gibson, W.~B.~Burton and D.~G.~Barnes,
  astro-ph/0110416.
  
  
\bibitem{Ackermann:2012mt}
  M.~Ackermann {\it et al.}   [Fermi-LAT Collaboration],
  arXiv:1207.6275 [astro-ph.HE].

\bibitem{FSW}
D.~Finkbeiner, M.~Su and C.~Weniger, to appear.
 
  
\end{thebibliography}
\end{document}